%% file: frontiers.tex
\documentclass[utf8]{FrontiersinHarvard} 

\usepackage{url,hyperref,microtype,subcaption}
\usepackage[onehalfspacing]{setspace}
\usepackage{upgreek}
\usepackage{booktabs}

\def\keyFont{\fontsize{8}{11}\helveticabold }
\def\firstAuthorLast{R. Rossini {et~al.}} 
\def\Authors{
R. Rossini\,$^{1,2,3*}$, 
G. Baldazzi\,$^{4,5}$, 
S. Banfi\,$^{6}$, 
M. Baruzzo\,$^{7}$, 
R. Benocci\,$^{6,8}$, 
R. Bertoni\,$^{6}$, 
M. Bonesini\,$^{6,9}$, 
S. Carsi\,$^{6,10}$, 
D. Cirrincione\,$^{7,11}$, 
M. Clemenza\,$^{6}$, 
L. Colace\,$^{12,13}$, 
A. de Bari\,$^{1,2}$, 
C. de Vecchi\,$^{2}$, 
E. Fasci\,$^{14,15}$, 
R. Gaigher\,$^{6}$, 
L. Gianfrani\,$^{14,15}$, 
A.D. Hillier\,$^{3}$, 
K. Ishida\,$^{3,16}$, 
P.J.C. King\,$^{3}$, 
J.S. Lord\,$^{3}$, 
R. Mazza\,$^{6}$, 
A. Menegolli\,$^{1,2}$, 
E. Mocchiutti\,$^{7}$, 
S. Monzani\,$^{7,11}$,
L. Moretti\,$^{14,15}$, 
C. Petroselli\,$^{6,10}$,
C. Pizzolotto\,$^{7}$, 
M.C. Prata\,$^{2}$, 
M. Pullia\,$^{2,17}$, 
L. Quintieri\,$^{3}$, 
R. Ramponi\,$^{18,19}$, 
M. Rossella\,$^{2}$, 
A. Sbrizzi\,$^{5}$, 
G. Toci\,$^{20}$, 
L. Tortora\,$^{13}$, 
E.S. Vallazza\,$^{6}$,
K. Yokoyama\,$^{3}$
and A. Vacchi\,$^{7,11}$}
\def\Address{%
$^{1}$Department of Physics, University of Pavia, Pavia, Italy \\
$^{2}$Sezione di Pavia, Istituto Nazionale di Fisica Nucleare (INFN), Pavia, Italy \\
$^{3}$ISIS Neutron and Muon Source, Science and Technology Facilities Council (STFC), Didcot, United Kingdom \\
$^{4}$Department of Physics "A. Righi", University of Bologna, Bologna, Italy \\
$^{5}$Sezione di Bologna, Istituto Nazionale di Fisica Nucleare (INFN), Bologna, Italy \\
$^{6}$Sezione di Milano-Bicocca, Istituto Nazionale di Fisica Nucleare (INFN), Milan, Italy \\
$^{7}$Sezione di Trieste, Istituto Nazionale di Fisica Nucleare (INFN), Trieste, Italy \\
$^{8}$Department of Earth and Environmental Sciences (DISAT), University of Milano-Bicocca, Milan, Italy \\
$^{9}$Department of Physics "G. Occhialini", University of Milano-Bicocca, Milan, Italy \\
$^{10}$Department of Science and High Technology, University of Insubria, Como, Italy \\
$^{11}$Department of Mathematics, Computer Science and Physics, University of Udine, Udine, Italy \\
$^{12}$Department of Engineering, University of Roma Tre, Rome, Italy \\
$^{13}$Sezione di Roma Tre, Istituto Nazionale di Fisica Nucleare (INFN), Rome, Italy \\
$^{14}$Department of Mathematics and Physics, University of Campania "Luigi Vanvitelli", Caserta, Italy \\
$^{15}$Sezione di Napoli, Istituto Nazionale di Fisica Nucleare (INFN), Naples, Italy \\
$^{16}$RIKEN Nishina Center, Saitama, Japan \\
$^{17}$Centro Nazionale di Adroterapia Oncologica (CNAO), Pavia, Italy \\
$^{18}$Sezione di Milano, Istituto Nazionale di Fisica Nucleare (INFN), Milan, Italy \\
$^{19}$Istituto di Fotonica e Nanotecnologie (IFN), Consiglio Nazionale delle Ricerche (CNR), Milan, Italy \\
$^{20}$Istituto di Nazionale di Ottica (INO), Consiglio Nazionale delle Ricerche (CNR), Florence, Italy 
}
\def\corrAuthor{Riccardo Rossini}

\def\corrEmail{riccardo.rossini@infn.it}

\begin{document}
\onecolumn
\firstpage{1}

\title[The muon beam monitor for the FAMU experiment]{The muon beam monitor for the FAMU experiment: design, simulation, test and operation} 

\author[\firstAuthorLast ]{\Authors} 
\address{} 
\correspondance{} 

\extraAuth{}

\maketitle

\begin{abstract}
\section{}
FAMU is an INFN-led muonic atom physics experiment based at the RIKEN-RAL muon facility at the ISIS Neutron and Muon Source (United Kingdom). The aim of FAMU is to measure the hyperfine splitting in muonic hydrogen to determine the value of the proton Zemach radius with accuracy better than 1\%.
The experiment has a scintillating-fibre hodoscope for beam monitoring and data normalisation. 
In order to carry out muon flux estimation, low-rate measurements were performed to extract the single-muon average deposited charge. 
Then, detector simulation in Geant4 and FLUKA allowed a thorough understanding of the single-muon response function, crucial for determining the muon flux. 
This work presents the design features of the FAMU beam monitor, along with the simulation and absolute calibration measurements in order to enable flux determination and enable data normalisation.

\tiny
 \keyFont{ \section{Keywords:} beam monitor, muon, beam calibration, single particle beam, muonic atom physics, detector simulation} 
\end{abstract}

\input{chapters/1intro}

\input{chapters/2hodos}
\input{chapters/3hodo4RAL}

\input{chapters/4analysis}
\input{chapters/5simulation}
\input{chapters/6results}
\input{chapters/7conclusion}

\section*{Acknowledgements}
The National Scientific Commission 3 (CSN3) of \textit{Istituto Nazionale di Fisica Nucleare} (INFN) is gratefully acknowledged as the funding agency of the FAMU experiment. 

The \textit{Science and Technology Facilities Council} (STFC) is gratefully acknowledged for the beamtime at the ISIS Neutron and Muon Source for the FAMU experiment (beamtime reference number RB2000022).

Computing resources were provided by INFN CNAF Tier1 (data analysis and Geant4 simulation) and STFC Scientific Computing Department's SCARF cluster (FLUKA simulation).

Authors would like to thank T. Schneider (CERN) for the cut and polishing of the 1 mm squared scintillating optical fibres used in this detector.

\section*{Conflict of Interest Statement}

The authors declare that the research was conducted in the absence of any commercial or financial relationships that could be construed as a potential conflict of interest.







\bibliographystyle{Frontiers-Harvard} 
\bibliography{test}

\end{document}

%% file: chapters/1intro.tex
\section{Introduction}\label{int}

\begin{figure}[t]
    \centering
    \includegraphics[width=0.6\linewidth]{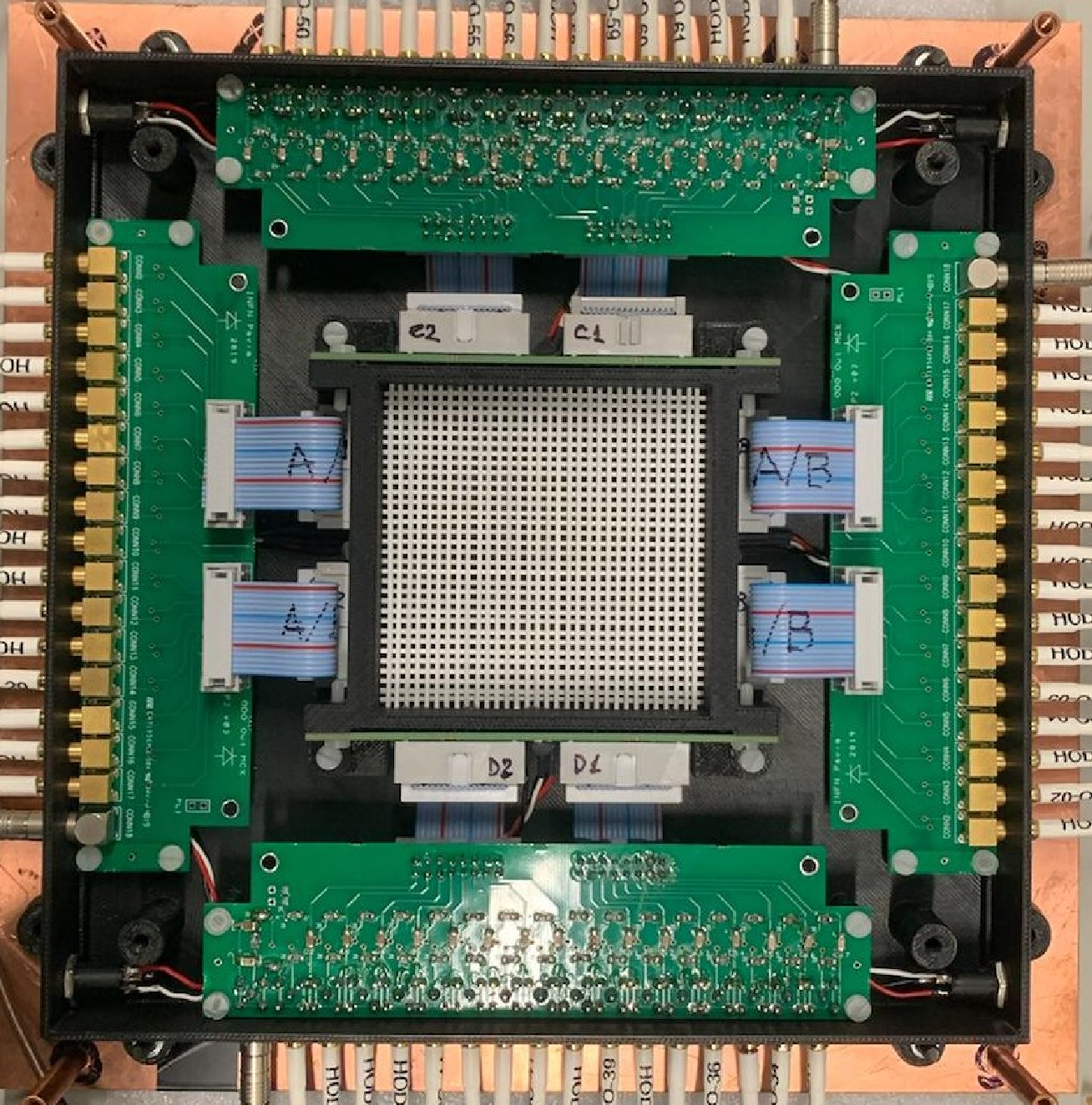}
    \caption{The hodoscope without its cover. The interspaced scintillating fibres are clearly visible in the middle (white). The internal PCBs hold the SiPMs ans they are connected by stripes to the external ones, holding the 64 MCX (signals) and 4 LEMO (bias supply) connectors.}
    \label{fig:hodo_open}
\end{figure}

The aim of the FAMU experiment (\cite{pizzolotto2020, Vacchi, rossiniFAMU:2023setup}) is to explore the magnetic structure of the proton through a measurement of the proton Zemach radius (\cite{Carlson, Antognini}).
The latter is extracted from a measurement of the hyperfine splitting energy of the muonic hydrogen ($\upmu$H) ground state.
The production of $\upmu$H atoms is performed by injecting a high-rate low-momentum pulsed muon beam into a pressurised gaseous target. The experiment is currently in operation at the RIKEN-RAL muon facility (\cite{matsuzaki2001, hillier2019}) at the ISIS Neutron and Muon Source (Didcot, UK). The observable of the experiment is the number of delayed muonic oxygen ($\upmu$O) X-rays resulting from the transfer of the muon from $\upmu$H to oxygen atoms. This is clearly dependent on the number of $\upmu$H atoms created, which is directly related to the incoming muon flux. As a consequence, having an accurate and efficient beam monitor, with minimal beam absorption, is a crucial point in the data normalisation. 

A beam hodoscope, composed of two planes of 32 scintillating fibres read-out by SiPMs, has been set-up for the experiment. The specific design of this detector, discussed in Section \ref{hod}, is the best match among the number of available channels (64), the detector area and its thickness. Other detector designs such as muon cameras (\cite{Lord}) were avoided in order to minimise the amount of material immersed in the beam, as the beam monitor is expected to stay in the beam for the duration of the experiment. Similar detectors, for higher rates and continuous beams, are being developed at other muon facilities such as PSI (\cite{papa:2015, papa:2019, dalmaso:2023}). The hodoscope serves both as a beam shape detector, to optimise beam centering and focusing, and as a flux-meter. The latter role of the detector is made possible thanks to the analyses reported in this work.

The estimated average negative muon flux with a momentum of 55 MeV/c is the order of $10^4$ muons per second (\cite{matsuzaki2001, hillier2019}). The beam is delivered in two 70 ns spills with an average repetition rate of 40 Hz (the synchrotron rate is 50 Hz, but one in five pulses is directed to the other target station). Therefore, during a spill, about 100 muons are delivered in 70 ns. Even though the system is based on a SiPM readout with fast signals ($\sim 20$ ns), it is clearly not possible to tell single-particle signals apart. For this reason, the detector measures the total deposited charge $Q_{tot}$, to be converted into muon flux using the result coming from calibration measurements (\cite{carbone:2015, bonesini2017, bonesini2018, rossini2023_1, rossini2023_2, rossiniHodoJINST}). 

Initially, data from cosmic muon calibrations combined with PDG $dE/dx$ results, to match the gap between energies around 4 GeV and the used beam momentum ($\sim 60$ MeV/c), were used to obtain an estimate of the muon flux vs. muon beam momentum (\cite{bonesini2017}), which compared well with previously published results (\cite{matsuzaki2001}). In this case, two 3 mm pitch hodoscopes (\textit{Hodo-2} and \textit{Hodo-3}) were used. 

Then, a 1 mm pitch hodoscope with adjacent fibres (\textit{Hodo-1}) was calibrated at the CNAO synchrotron in Pavia (Italy) with a low-rate proton beam with energy loss $dE/dx$ comparable to FAMU muons (\cite{rossiniHodoJINST}). A proton beam of kinetic energy 150 MeV was tuned to allow single particle events and directed against the hodoscope for testing.

The latest FAMU hodoscope (\textit{Hodo-4}), i.e. the position-sensitive muon beam monitor detector, is here thoroughly discussed. The design of the detector, discussed in Section \ref{hod}, fits best with the number of available channels (64), and the required thickness, active area and space resolution. It is composed of 1 mm fibres, spaced by 1 mm. The detector full simulation and tests are presented in this work. Calibration measurements were carried out in the FAMU setup, exploiting the RIKEN-RAL Port1 muon beam with a modified configuration in order to obtain a low-rate muon beam, as later explained in Section \ref{ral}. This has been crucial in order to disclose single-particle signals. In addition, the detector has been simulated in Geant4 (\cite{GEANT}) and Flair-FLUKA (\cite{FLUKA1, FLUKA2, FLAIR}) in order to understand its response and extract crucial parameters and information for its calibration.

The equation to extract the muon rate from the hodoscope reading is the following: 
\begin{equation}\label{eq:flux}
    \varphi_\mu = 
    \frac{r}{\left( W_2 + \frac{W_1}{\eta} \right) Q_\mu}Q_{tot} =: k Q_{tot} 
\end{equation}
where $r=40$ Hz is the beam repetition rate, $Q_{tot}$ is the total deposited charge during a full-rate beam spill, $Q_\mu$ is the average charge deposited by a 55 MeV/c muon interacting with both planes of fibres, $Q_\mu/\eta$ for muons interacting with one fibre only, and $W_{1/2}$ are the fractions of muons interacting with 1 or 2 fibres, respectively. The charge deposited in one fibre is written as $Q_\mu/\eta$ because this value is not directly measured and $\eta$ is calculated from the simulation. In particular, $W_2$ is extracted from the simulation, while $Q_\mu$ is determined using low-rate data. The main aim of this work, i.e. the \textit{calibration} of the FAMU beam monitor, is to compute the value of the calibration factor $k=\frac r{(W_2 + W_1/\eta)Q_\mu}$. This work provides a general method to calibrate other fibre-based hodoscopes to be used as charged particle beam monitors. 

Simulation, measurements and analysis techniques are presented in Sections \ref{ral}, \ref{ana} and \ref{sim}, whereas the results are extensively presented and discussed in Section \ref{res}. Eventually, the value of $k$ is computed and a test estimation of flux during full-rate beam is shown.

%% file: chapters/2hodos.tex
\section{Design of the 1 mm hodoscope (Hodo-4)}\label{hod}

\begin{figure}[t]
    \centering
    \includegraphics[width=\linewidth]{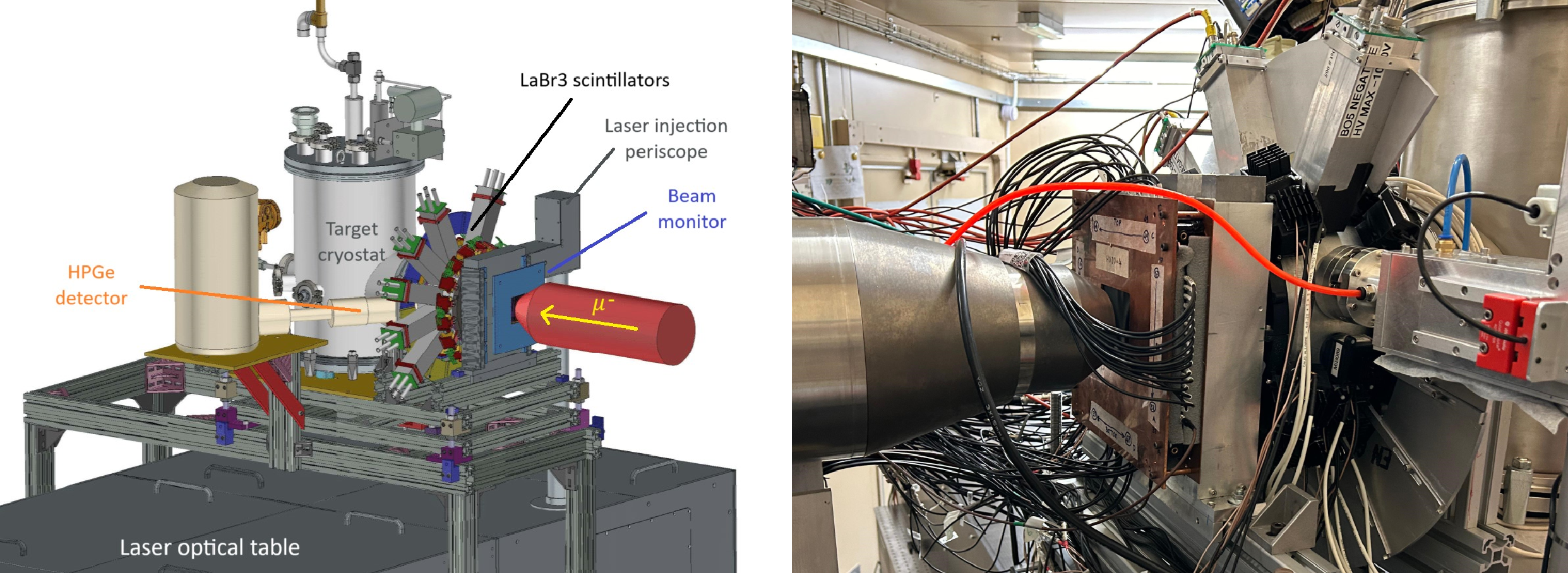}
    \caption{FAMU in the 2023 setup. The picture on the left shows a CAD scheme of the target and detector system in the 2023 setup: the hodoscope (beam monitor) is positioned just after the muon beamline collimator, whereas all other detectors (LaBr$_3$ and HPGe) are positioned around the gas target. On the right is shown a picture of the setup, where the hodoscope (copper-shielded detector) is in close-up. Figures taken from \cite{rossiniHodoJINST}.}
    \label{fig:FAMUcadpic}
\end{figure}

A $32 \times 32$ channels (XY configuration) beam monitor has been set up for the FAMU experiment by INFN Milano-Bicocca and INFN Pavia. The hodoscope consists of two crossed planes of 32 single-clad Saint-Gobain/Luxium BCF-12 polystyrene scintillating fibres each. Each fibre is squared, with a pitch of 1 mm, and each fibre is covered with a nominal $15 \ \upmu$m-thick layer of TiO$_2$-based ExtraMural Absorber (EMA, or coating) to avoid inter-fibre optical cross-talk. 

Previous hodoscope versions (\cite{carbone:2015, bonesini2018, rossini2023_1, rossini2023_2, rossiniHodoJINST}) had either too much material immersed in the beam ($32 \times 32$ fibres with 3 mm-pitch, i.e. active area of $9.6 \times 9.6$ cm$^2$ and total thickness of $6$ mm) or too small active area ($32 \times 32$ fibres with 1 mm-pitch, i.e. active area of $3.2 \times 3.2$ cm$^2$ and total thickness of $2$ mm). 
The key point of the detector described in this work is having a spacing of 1 mm between adjacent fibres, as one can see in Figure \ref{fig:hodo_open}. This allows to have a $6.4\times6.4$ cm$^2$ detector area, despite keeping the thickness below 2 mm. However, this hodoscope has inhomogeneous volume, which slightly complicates its response function to the muon beam, as further discussed in Sections \ref{sim} and \ref{ana}. In fact, each muon can interact with 0, 1 or 2 fibres depending on its position of interaction in the XY plane. The geometric features of this model compared to the previous ones are resumed in Table \ref{tab:hodos}.

This model of the hodoscope has been considered the best compromise among all features and it has therefore been installed in the final FAMU setup for 2023 and 2024 runs. As a consequence, carrying out single-particle calibration was crucial for its operation in the FAMU experiment. The position of the hodoscope in the FAMU setup is shown in Figure \ref{fig:FAMUcadpic}.

Each fibre is read-out by a $1 \times 1$ mm$^2$ Hamamatsu S12571-050P SiPM (cell size 50 $\upmu$m) on one side. SiPMs are supplied with a positive bias of $66.64 \pm 0.12$ V (finely tuned for each group of 16 SiPMs to optimise the uniformity of the detector response)\footnote{the operational voltage values for the three PCBs (each one holding 16 SiPMs) are: $66.45 \pm 0.03$ V, $66.65 \pm 0.01$ V, $66.69 \pm 0.01$ V and $66.78 \pm 0.02$ V.}. SiPM signals are fanned out through 4 m cables with MCX connectors and digitised using two CAEN V1742 (32 channels each). The trigger is supplied to all digitisers (including these ones) through the FAMU system, which provides beam trigger coming from the synchrotron with a rate of 50 Hz.

\begin{table}[ht]
    \caption{Comparison among the various models of $32 \times 32$ hodoscope developed for the FAMU experiment. Hodoscopes with 1 mm fibres have only 2 mm thickness but small active area, whereas those with 3 mm fibres have larger area, but they are 6 mm thick. The model described in this work and finally installed in FAMU has 1 mm fibres interspaced by 1 mm, allowing a mid-size active area without compromising the detector thickness.\newline}
    \centering
    \begin{tabular}{c|c|c|c}
        \toprule
        \textbf{Hodoscope} & \textbf{Fibre pitch} & \textbf{Thickness} & \textbf{Active area} \\
        \midrule
        Hodo-1    & 1 mm & 2 mm & $3.2\times3.2$ cm$^2$\\
        Hodo-2/3    & 3 mm & 6 mm & $9.6\times9.6$ cm$^2$\\
        Hodo-4   & 1 mm & 0-2 mm & $6.4\times6.4$ cm$^2$\\
        \bottomrule
    \end{tabular}
    \label{tab:hodos}
\end{table}

%% file: chapters/3hodo4RAL.tex
\section{Hodoscope measurements at RIKEN-RAL Port1}\label{ral}


The FAMU experiment is installed at the RIKEN-RAL Port1 muon beamline at the ISIS Neutron and Muon Source in Didcot, United Kingdom. The experiment consists of a pressurised cryostat holding $ \sim 7.5$ bar of hydrogen-oxygen mixture at a temperature of about 90 K. The gas chamber is the target of the muon beam, with the aim of forming muonic hydrogen atoms. 

\begin{figure}[t]
    \centering
    \includegraphics[width=0.6\linewidth]{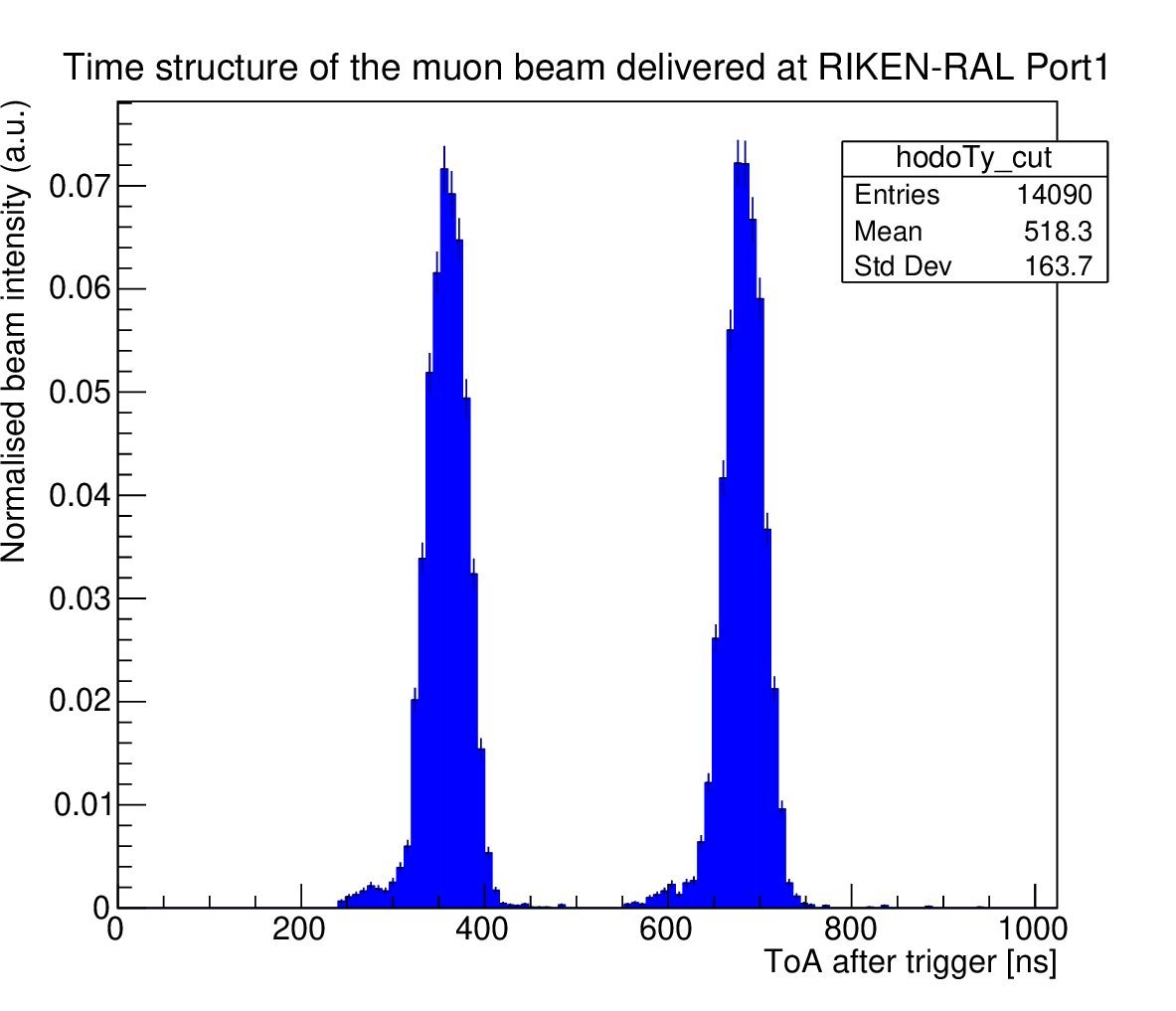}
    \caption{Time structure of the muon beam for each trigger. This normalised histogram has been measured by the FAMU hodoscope at low rate, but the time structure holds for high rate too.}
    \label{fig:ToA}
\end{figure}

The ISIS synchrotron accelerates protons with an energy of 800 MeV, with a pulse rate of 50 Hz. Four consecutive pulses are sent to Target Station 1 (TS1), and one is sent to Target Station 2 (TS2). The graphite target for muon beamlines is located in the beampipe connecting the synchrotron to TS1. Hence, it receives protons and produces muon pulses with the same rate as TS1 ($r = 40$ pulses per second). Negative pions are directed in the RIKEN beampipe, where they decay to negative muons and are delivered to the four RIKEN-RAL Ports alternatively. RIKEN-RAL Port1 is currently dedicated to the FAMU experiment. The beam time structure, which is the same for high and low rate measurements, is shown in Figure \ref{fig:ToA} as measured by the FAMU hodoscope in a low-rate test beam. Each beam pulse consists of two 70 ns spills separated by 320 ns. 

\begin{figure}[t]
    \centering
    \includegraphics[width=0.9\linewidth]{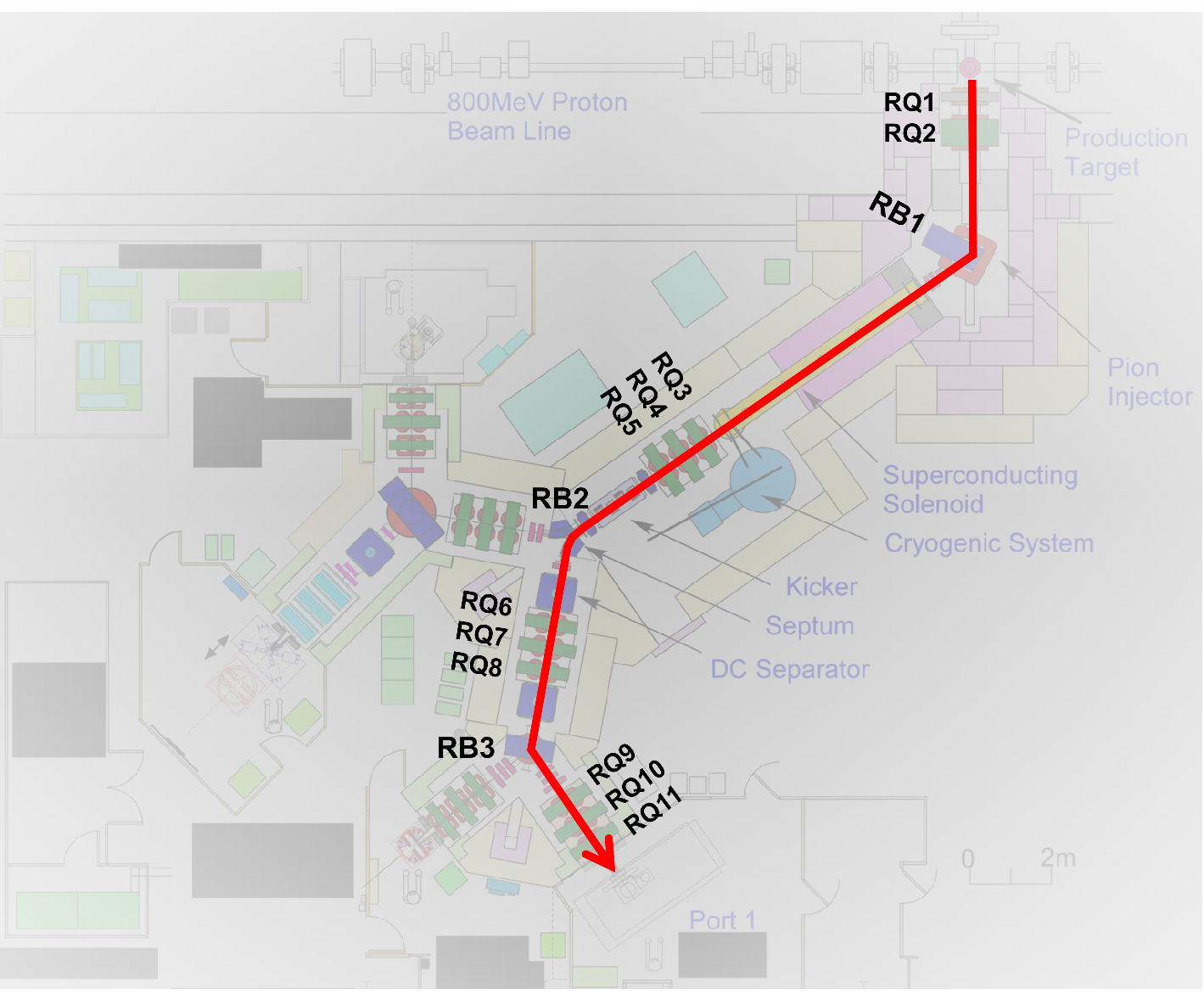}
    \caption{Path followed by the negative muons (red arrow) from the graphite target to Port1, where the FAMU target and detectors are located. The quadrupole (RQ) an bending magnets (RB) implied in the delivery of muons to FAMU are labelled. Adapted from \url{https://www.isis.stfc.ac.uk/}.}
    \label{fig:RIKEN}
\end{figure}

\begin{figure}[t]
    \centering
    \includegraphics[width=0.6\linewidth]{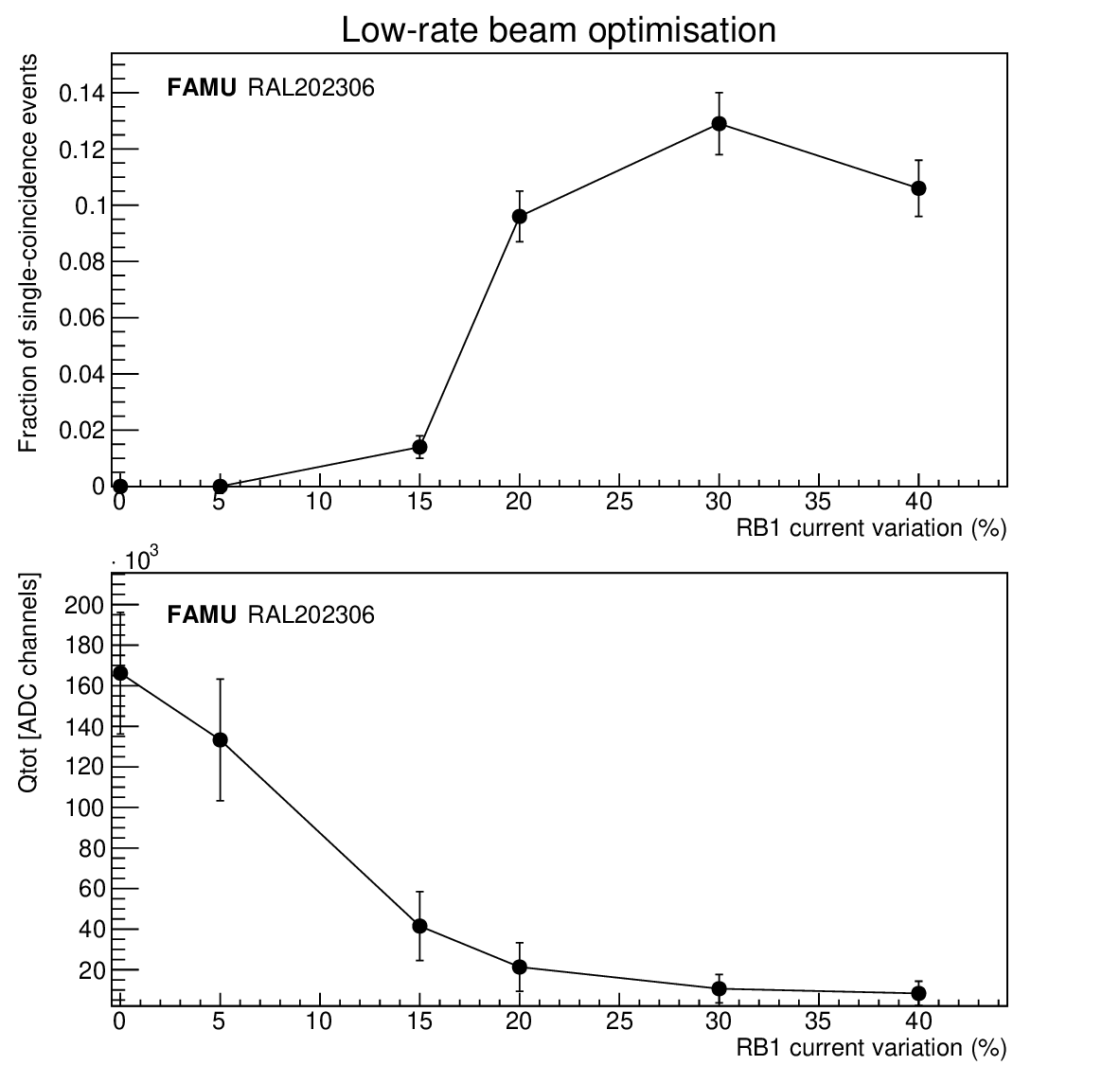}
    \caption{Optimisation of the RB1 current to obtain the low-rate configuration. The upper panel shows the fraction of events selected as single-particle events as a function of the RB1 current displacement, while the lower one shows the related decrease of $Q_{tot}$. Working at RB1$+40\%$ guarantees low-rate conditions.}
    \label{fig:lowrate}
\end{figure}

    \subsection{High-rate measurements at RIKEN-RAL}
    The beam is generally set to work at the highest available rate. At the momentum value used in FAMU, the average rate is the order of $10^4$ muons/s. The quadrupole and bending magnet configuration has been optimised to deliver the best beam rate and geometry that suits for the experiment during the FAMU beam commissioning in July 2023 (4 beam days, dataset \textit{RAL202303}). 
    
    After that, two FAMU data taking runs have been carried out in October 2023 (6 beam days, \textit{RAL202305}) and December 2023 (12 beam days, \textit{RAL202306}).
    
    \subsection{Low-rate measurements at RIKEN-RAL}
    After carrying out some tests with protons at the CNAO synchrotron in Pavia (\cite{rossiniHodoJINST}), it has been decided to characterise the current hodoscope directly on the FAMU setup as its particular design was expected to be more sensitive to beam geometry changes.

    In order to obtain a single particle beam, the  currents of some quadrupole and bending magnets were de-tuned in order to minimise the amount of pions directed in the beampipe. In particular, the first two quadrupoles (RQ1 and RQ2) were turned off in order to widen the pion bunch, and the first bending magnet (RB1) was de-tuned in order to direct the beam halo, and not its central part, into the beampipe. See Figure \ref{fig:RIKEN} for a detailed map of the path followed by the beam from the target to Port1, including the magnets encountered. The choice of which magnets had to be tuned was done in order not to compromise the beam optics, which would result in not delivering the beam to Port1. The shut-down of the two quadrupoles resulted in a $\sim$90\% beam intensity drop. The optimisation of the bending magnet current was carried out progressively in order to make sure that the rate would be as low as required. Figure \ref{fig:lowrate} shows the effect of the progressive variation of the RB1 current out if its optimal value. The muon current (proportional to $Q_{tot}$) decreases, whereas the number of events marked as single coincidences increases, reaches a maximum and starts to decrease. This latter behaviour means that the muon flux is so low it allows events with only one coincidence, i.e. single-muon spills.
    
    The low-rate data acquisition \textit{RAL202306} consisted of a couple of hours of beam optimisation and final $50k$-events measurements with and without beam.

\begin{figure*}[t!]
    \centering
    \includegraphics[width=0.95\linewidth]{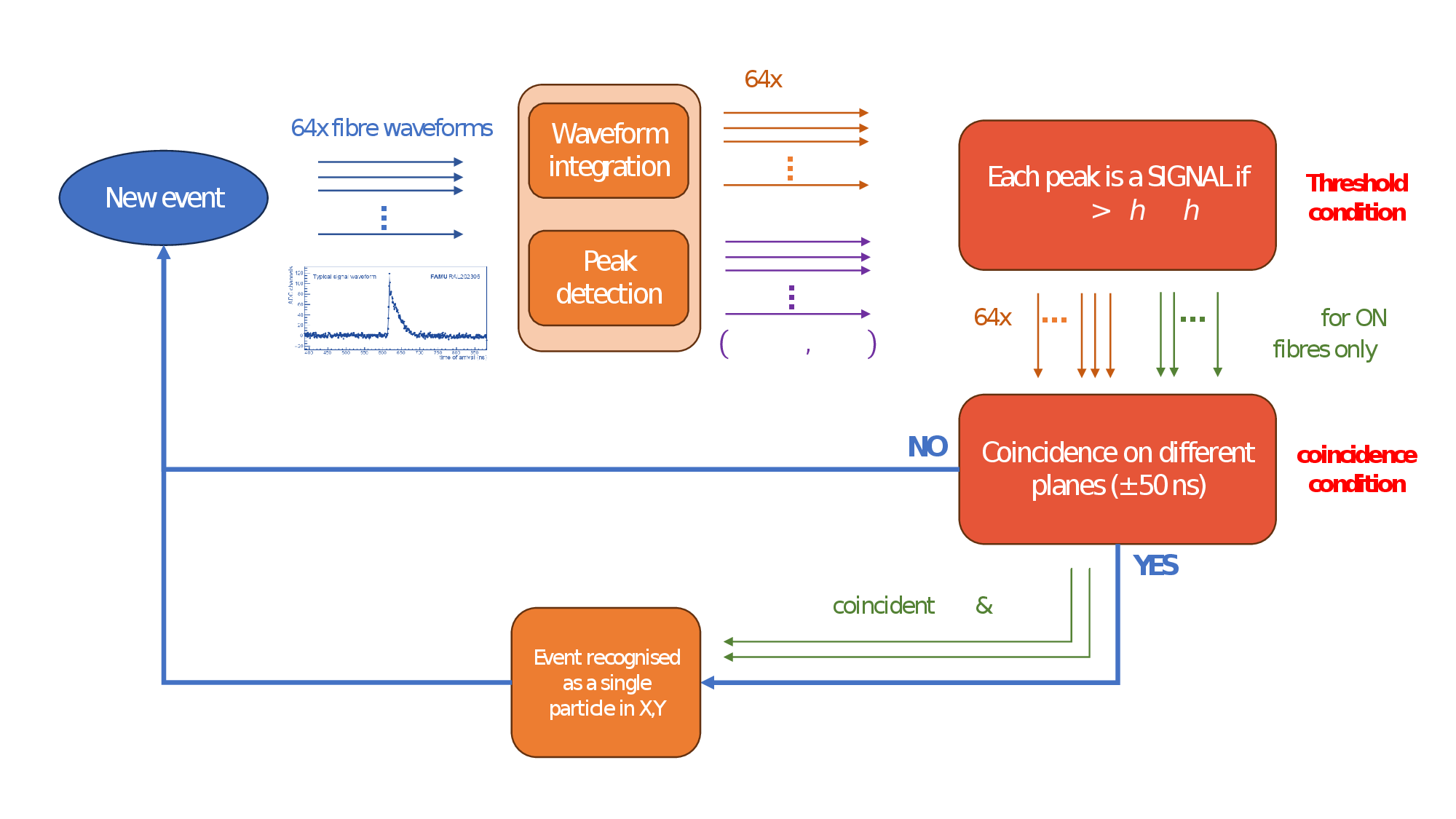}
    \caption{Scheme of the coincidence data processing workflow, described in Section \ref{ana}, to select events with a single muon interacting with both detector planes (\textit{2-hit} data).}
    \label{fig:analysis}
\end{figure*}

%% file: chapters/4analysis.tex
\section{Data analysis}\label{ana}

The data analysis technique, described in Figure \ref{fig:analysis}, is based on imposing single coincidences between each plane of the detector. 

For each beam trigger, the FAMU DAQ system opens an acquisition window and digitises the signals coming from each hodscope channel with a rate of 1 GS/s. The resulting data packet is called an event. In the previous analysis procedure, extensively described and tested in \cite{rossiniHodoJINST}, an event was considered a single-particle hit if only one fibre per plane had integrated charge over a given threshold. This method has been considered valid as the hodoscopes had no interspacing and the measurements were particle-triggered. This means that most muons arrived at the same time (given the fixed pre-trigger window) and passed by one fibre per plane. However, in this case the trigger comes from the synchrotron and the full-rate beam shape is complex (every synchrotron trigger corresponds to two $\sim 70$ ns spills separated by a $\sim 320$ ns gap). The low-rate time structure of the beam is the same: even though the beam is tuned to allow single-particle events, they might come from either the first or the second spill, and some events might have more than one muon. In addition, the spacing between adjacent fibres makes it less probable to have muons hitting one fibre per plane, as better discussed and quantified in Section \ref{sim}. For this reason, it's been decided to use a time coincidence-based approach.

During data processing, for each event and for each hodoscope fibre $j$, the hodoscope low-rate data processing system retrieves the 64 waveforms and looks for peaks, returning the total integrated charge $Q_j$ and, for every peak $k$, the time-of-arrival $t_j^k$ and the pulse height $\text{PH}_j^k$. At this point, the coincidence is imposed, with a tolerance of 50 ns (small enough to exclude particles coming from two different spills), for hodoscope peaks having $\text{PH}_j^k$ over a certain threshold to be determined. Events having only one coincidence are selected as single-particle events and therefore used for the hodoscope characterisation.

The value of total deposited energy for every event is the sum on all fibres of the integrated charge $Q_{tot} = \sum_{j=1}^{64} Q_j$. This holds for both low- and high-rate measurements. As one can see in Section \ref{res}, the shape of the $Q_{tot}$ histogram for low-rate measurements is asymmetric. After exploring some possibilities (combination of Landau and Gauss profiles), it has been decided to fit this histogram with the convolution of a Gaussian and a decreasing exponential profile, i.e.:
\begin{align}\label{eq:convolution}
    F(x) &= A \int_{-\infty}^{x}dt \ e^{-t/\tau} \ G_{\mu,\sigma}(x-t) \\
    &= C+A\exp \left[ -\frac{x-\mu}{\tau} + \frac{\sigma^2}{2\tau^2}\right] \left( 1+ \text{erf} \left( \frac{x-\mu-\sigma^2/\tau}{\sqrt{2}\sigma} \right) \right) \nonumber
\end{align}
having 5 free parameters: additive constant $C$, amplitude $A$, Gaussian mean $\mu$, Gaussian sigma $\sigma$ and exponential decay constant $\tau$. The fit boundaries are chosen by looking for optimal and stable reduced $\chi^2$. The maximum, which corresponds to the estimate for $Q_\mu$, has no known analytical expression. As a consequence, it has been determined on the fit function through numerical Brent method, using ROOT\cite{ROOT} method TF1::GetMaximumX(). The uncertainty on $Q_\mu$ is obtained by variations of the fit boundaries around the optimum. This is done recursively in order to select a region in the two fitting boundaries in which $\chi^2/NDF < 1.3$. The variation of $Q_\mu$ in this region is then used as an estimation for its uncertainty.

\begin{figure}[t]
    \centering
    \includegraphics[width=0.75\linewidth]{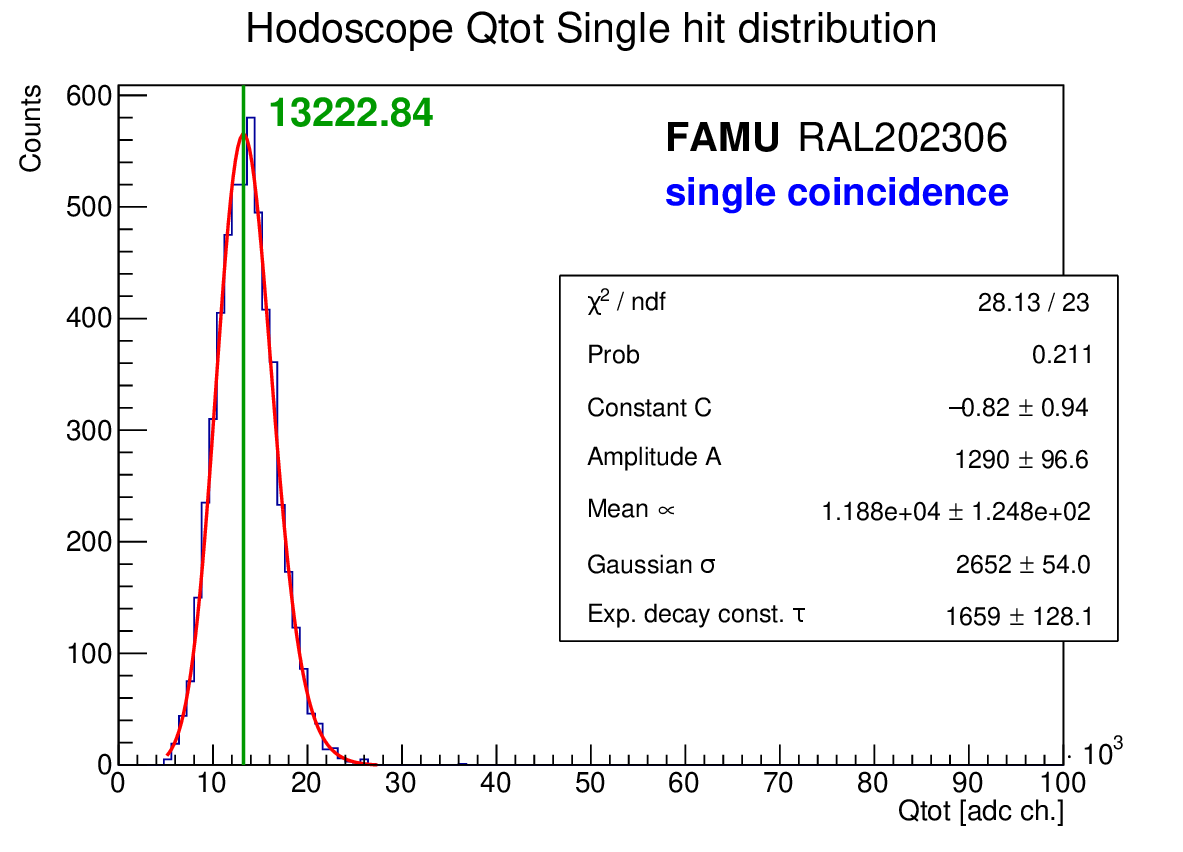}
    \caption{Determination of $Q_\mu$ from the fit of the low-rate $Q_{tot}$ histogram. The fitting function is the distribution in Eq. \ref{eq:convolution} and the maximum is obtained with numerical Brent method.}
    \label{fig:Qmu}
\end{figure}

The value obtained by analysing the data taken at RIKEN-RAL with low-rate muons is $Q_\mu = (14080 \pm 50)$ ADC channels. The uncertainty is taken by varying the fitting boundaries and imposing $\chi^2/NDF < 1.3$. The histogram and fit are shown in Figure \ref{fig:Qmu}. This is the most probable value of deposited charge by 55 MeV/c muons interacting with two scintillating fibres, one per plane. As discussed, the muon flux with this geometry can not be simply obtained as $Q_{tot}/Q_\mu$, as most muons don't interact with two fibres. However, the fraction of muons interacting with one fibre per plane ($W_2$) and with one fibre only ($W_1$) are mostly geometric and have to be extracted from the simulation. Therefore, the flux can be estimated with Eq. \ref{eq:flux}, i.e. weighing $Q_{tot}/Q_\mu$ by a factor $1/(W_2 + W_1/\eta)$, which is obtained from the simulation in Section \ref{sim}.


%% file: chapters/5simulation.tex
\section{Hodoscope simulation}\label{sim}

In order to understand the energy loss of muons in the detector, and therefore its theoretical response function, the hodoscope has been simulated using the Geant4 (\cite{GEANT}) toolkit. The geometry consists of the fibres, coatings and entrance windows as described in Section \ref{hod}, i.e.: each fibre (polystyrene, 1 mm pitch, 6.4 cm length) is coated with a 15 $\upmu$m layer of TiO$_2$ and positioned in a 32-fibre plane with 1 mm inter-spacing between adjacent fibres (measured coating-to-coating); two planes are juxtaposed with crossing fibre direction, separated from the world volume with a 0.1 mm-thick PVC window. 

The muon beam simulated for this work is a 55 MeV/c negative muon beam with 2-dimensional gaussian shape, with $\sigma_X = (8.15 \pm 0.02)$ mm and $\sigma_Y = (10.354 \pm 0.012)$ mm. The reproduces the beam configuration optimised for the experiment, as measured with the hodoscope during 12 hours of full-rate data acquisition. 

In order to obtain an uncertainty budget, all simulations were repeated with gaussian dispersion of momentum ($\sigma_p/p = 10$\%), variable beam size within the $\sigma_X$-$\sigma_Y$ uncertainties, and variable coating thickness, considering a 5 $\upmu$m coating thickness tolerance. This resulted in a geometric systematic uncertainty, which was added to the uncertainty budget as an independent contribution. 

\begin{figure}[ht]
    \centering
    \includegraphics[width=0.75\linewidth]{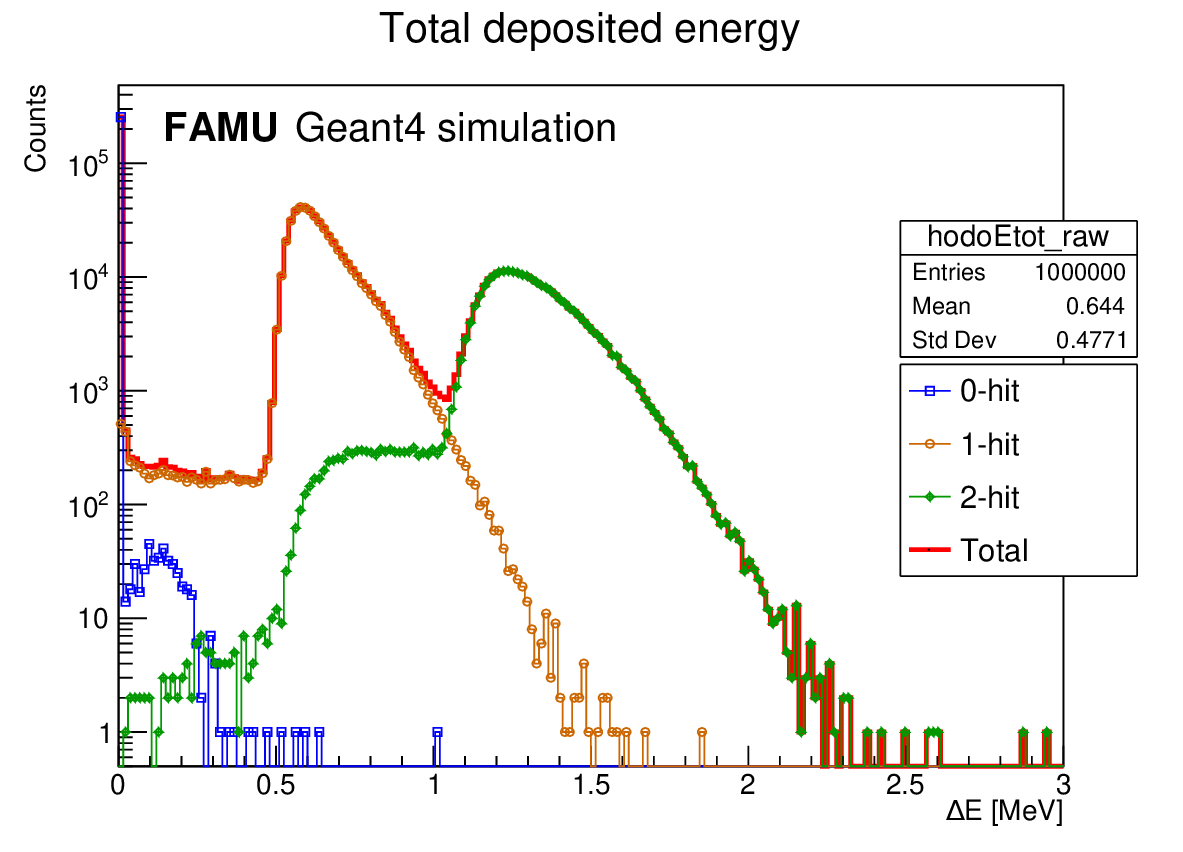}
    \caption{Geant4 simulation ($10^6$ events): energy deposited in scintillating fibres by muons interacting with 0 (blue squares), 1 (orange circles) and 2 (green diamonds) fibres. The total contribution (thick red line) is the expected response function of the detector.}
    \label{fig:Esim}
\end{figure}

\begin{figure}[t]
    \centering
    \includegraphics[width=0.6\linewidth]{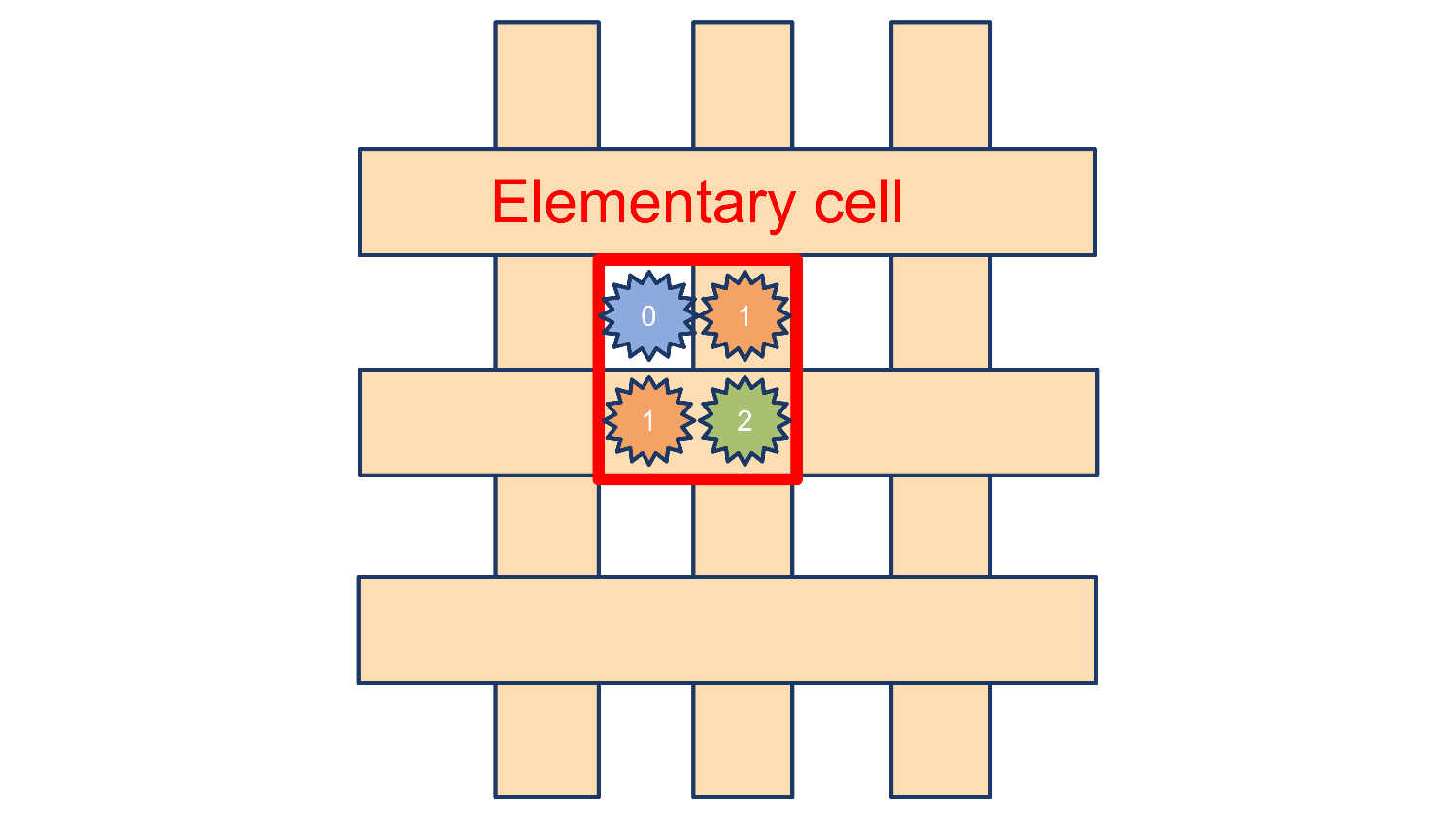}
    \caption{Graphical representation of the hodoscope subdivision in elementary cells (red). As one can derive from the scheme, assuming uniform flux, muons can interact with either 0 (blue), 1 (orange) or 2 fibres (green) with probabilities $\sim$ 25\%, 50\% and 25\%, respectively. As a consequence, the heuristic values to be compared to the simulation are $W_2 \sim 0.25$ and $W_1 \sim 0.5$.}
    \label{fig:Cell}
\end{figure}

All primaries have been tracked and assigned flags depending on whether they passed by front and back plane fibres. In fact, given the geometry of the hodoscope, muons can pass by 0, 1 or 2 fibres. The contribution of these muons are plotted separately and also jointly in Figure \ref{fig:Esim}. In the 0-hit case, the energy deposit different from zero in some events is caused by secondary particles (\textit{e.g.} delta rays generated in the coating and decay electrons) interacting with the fibres. The probability of hitting 0, 1 or 2 fibres is about 25\%, 50\% and 25\%, respectively, as one can derive from geometrical considerations from Figure \ref{fig:Cell}. However, the exact values of $W_2$ and $W_1$ depend on beam geometry, scattering processes and coating thickness, as a consequence they have to be extracted from the simulation. $W_1$ is equal to the fraction of particles interacting with only 1 fibre, while $W_2$ is the fraction of muons passing by two fibres, one per plane. 


In the Geant4 simulation, $10^6$ negative muons with momentum 55 MeV/c were launched. The beam geometry is the one extracted from 12 hours of full-rate hodoscope measurements. As expected, about $\frac 1 4$ of the muons were marked as passing by two fibres, with a statistical counting uncertainty of around 0.2\%. The uncertainty budget was completed by repeating the simulation with variations in the fibre pitch within its tolerance (30 $\upmu$m), the coating thickness within 5 $\upmu$m, the beam momentum within 10\% and the beam shape within the measured uncertainty. Other effects such as small fibre misplacements are expected to be averaged and cancelled out due to the beam spot size. The total contribution, which is dominated by the uncertainty on the coating thickness, is about 1.8\% on the number of muons passing by 2 fibres. The final estimate for the double-hit fraction is $W_2 = (24.9 \pm 0.4)\%$. This value is consistent with the heuristic value of 25\% estimated from the geometry of the detector elementary cells.
Similarly, the single-hit fraction estimate is $W_1 = (49.61 \pm 0.09)\%$.

In addition, the ratio $\eta$ between the double-hit and the single-hit mean deposited energy had to be computed. In fact, the data selection described in Section \ref{ana} enables only the inclusion of double-hit events. As a consequence, the value of $Q_\mu$ calculated from the measurements is only the mean charge deposited by particles hitting two fibres. In principle, with a local linear approximation of the energy-loss curve, one could assume that the energy released by single-hitting muons is $Q_\mu/2$ (i.e. that $\eta=2$, but this has to be verified, as the linear approximation might not hold. To do so, the single- and double-hit simulated spectra (see Figure \ref{fig:Esim}) have been fitted with Equation \ref{eq:convolution}. The fit stability was tested and used to determine the uncertainties as in the case of data coming from low rate measurements (see Section \ref{ana}), along with parameter variation. The estimated values of deposited energy are $E_2 = (1.23 \pm 0.06)$ MeV for the 2-hit and $E_1 = (0.58 \pm 0.03)$ MeV for the 1-hit. The resulting beam momentum straggling, which comes from the sparse detector geometry, is comparable to the momentum bite of the incoming beam ($dp/p \sim 4$\% (\cite{matsuzaki2001})) and small compared to the stopping range in the FAMU apparatus. As a consequence, the presence of the hodoscope doesn't spoil the FAMU data. 

The estimate for the ratio between the double-hit and the single-hit mean deposited energy is $\eta = E_2/E_1 = 2.11 \pm 0.05$. 

\begin{figure}[ht]
    \centering
    \includegraphics[width=0.75\linewidth]{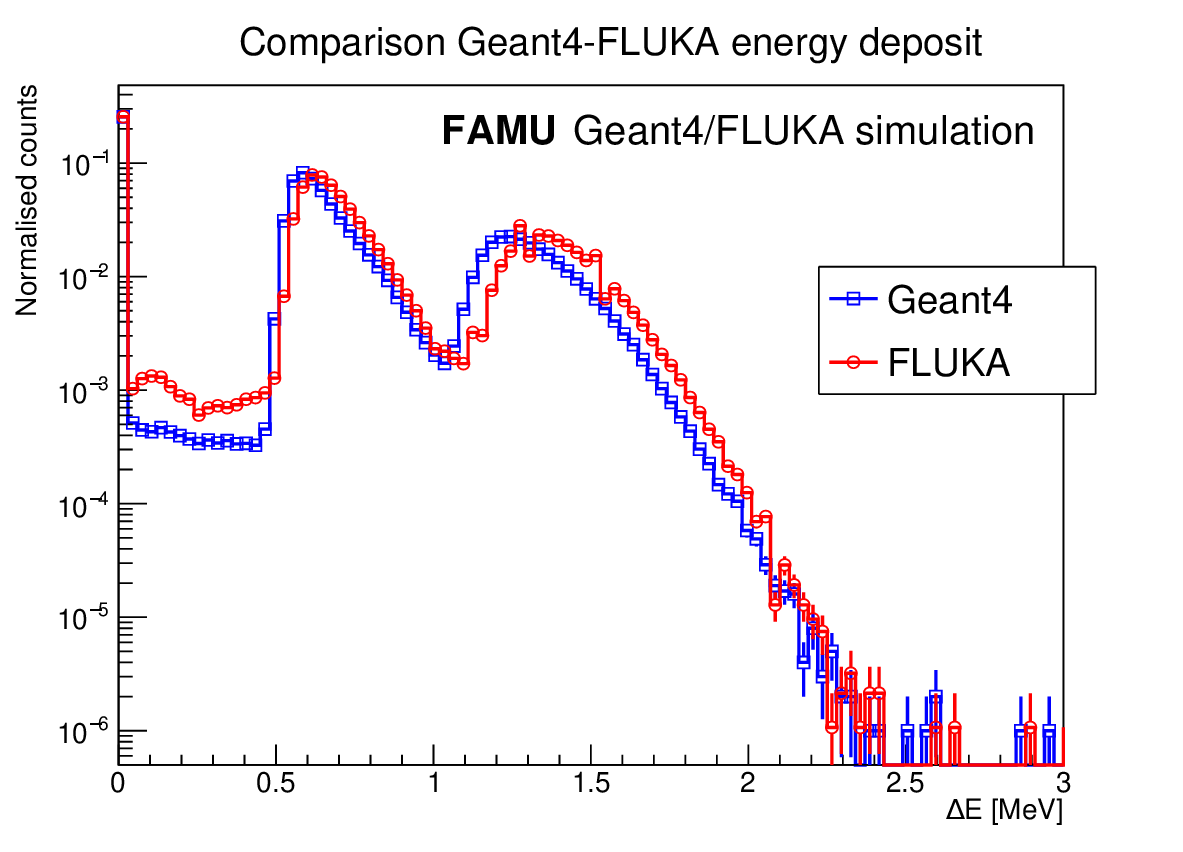}
    \caption{Comparison of Geant4 (blue squares) and FLUKA (red circles) simulated energy deposited in hodoscope fibres ($10^6$ events). The two energy peaks differ by 6-7\%, which is considered good qualitative accordance between the two approaches. Fine differences between the spectra are currently being investigated.}
    \label{fig:G4vsFLUKA}
\end{figure}

In parallel, an independent simulation based on the FLUKA-CERN (\cite{FLUKA1, FLUKA2}) toolkit was also developed using the Flair interface (\cite{FLAIR}) for comparison (reported just as FLUKA in this work, for simplicity). The FLUKA simulation has been modeled to match the exact geometry and beam characteristics with the one in Geant4. The geometric factors determined with the FLUKA simulation with the same number of events ($10^6$) are: $W^\mathtt{FLUKA}_2 = (25.17 \pm 0.06)\%$ and $W^\mathtt{FLUKA}_1 = (49.94 \pm 0.07)\%$. The uncertainties on the FLUKA predictions are underestimated as they only comprise the statistical component. Both values differ by less than 3 standard deviations from the values estimated in Geant4. The qualitative comparison between the FLUKA and Geant4 histograms of the energy deposited in the hodoscope active volumes is shown in Figure \ref{fig:G4vsFLUKA}. The deposited energy in 1-hit and 2-hit cases, obtained from fitting with Equation \ref{eq:convolution}, are $E^\mathtt{FLUKA}_2 = (1.32 \pm 0.03)$ MeV and $E^\mathtt{FLUKA}_1 = (0.62 \pm 0.02)$ MeV. These energy deposits differ by about 7\% and 6\%, respectively, from the Geant4 values. Such a difference is generally considered a sign of accordance between the results retrieved from two independent codes. Their ratio is $\eta^\mathtt{FLUKA} = 2.13 \pm 0.08$, which is consistent with the value extracted from the Geant4 simulation. 

The FLUKA simulation was also used to estimate effective backscattering, \textit{i.e.} the fraction of particles interacting with one fibre in the first plane, one in the second plane and then one back in the first plane, as a result of backscattering. Ideally, this value should be minimal for better hodoscope accuracy. The effective backscattering rate is $(0.013 \pm 0.004)$ \textperthousand, negligible.

\begin{table*}[t]
    \caption{Comparison between parameters estimated by the Geant4 and FLUKA simulation with $10^6$ events simulated. The uncertainty balances are obtained through parameter variation, whereas those marked with "stat" are statistical only. The values of $W_2$ are consistent with each other, while the values of $W_1$ are qualitatively comparable and differ by less than 3 standard deviations (t-Student test). The estimates of $E_2$, $E_1$  and $\eta$ are consistent between the two independent simulation toolkits.\newline}
    \centering
    \begin{tabular}{l|l|l|l|l|l}
        \toprule
        \textbf{Toolkit} & $\mathbf{W_2}$ & $\mathbf{W_1}$ & $\mathbf{E_2}$ (MeV) & $\mathbf{E_1}$ (MeV) & $\boldsymbol{\eta} = E_2/E_1$  \\
        \midrule
        Geant4    & $24.9 \pm 0.4$ & $49.61 \pm 0.09$ & $1.23 \pm 0.06$ & $0.58 \pm 0.03$ & $2.11 \pm 0.05$ \\
        FLUKA  &  $25.17 \pm (0.06)_\text{stat}$ & $49.94 \pm (0.07)_\text{stat}$ & $1.32 \pm 0.03$ & $0.62 \pm 0.02$ & $2.13 \pm 0.08$ \\
        \bottomrule
    \end{tabular}
    \label{tab:G4vsFLUKA}
\end{table*}

All the useful parameters extracted from Geant4 and FLUKA simulations are reported in Table \ref{tab:G4vsFLUKA}. The coefficients for Equation \ref{eq:flux} used in this work for flux estimation are taken from the Geant4 simulation, since the applied physics lists and transport thresholds had been already tuned and validated for the FAMU experiment. 

\begin{figure}[ht]
    \centering
    \includegraphics[width=\linewidth]{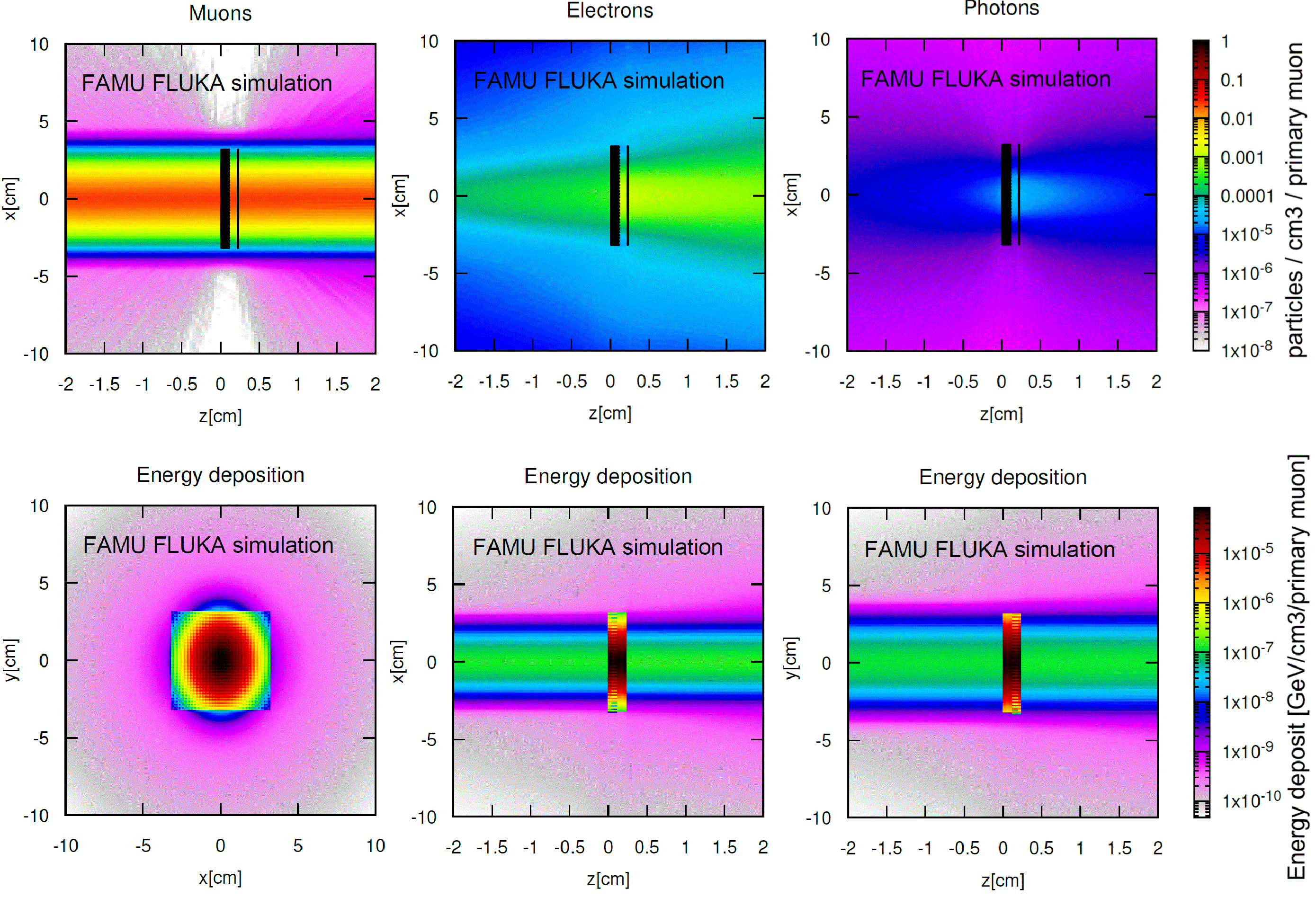}
    \caption{2D distributions extracted from the Flair interface to the FLUKA simulation. The muon beam runs along the $z$ axis in the positive direction (left to right). First row shows the density of muons/electrons/photons in the path followed by the muons, with the same color scale for all plots. The second row shows the space distribution of energy deposit.}
    \label{fig:2D-FLUKA}
\end{figure}

Figure \ref{fig:2D-FLUKA} shows some 2D distributions obtained using the Flair interface to FLUKA-CERN. In particular, the first row shows the probability density for the particles present in the simulation (primary muons, decay electron and photons/electrons resulting from elastic and inelastic processes). The second row shows the space distribution of the energy deposition, which is higher in the detector fibres than in air, reflects the gaussian shape of the beam on the $xy$ plane. Muon inelastic processes are negligible for the sake of the experiment. The main mechanism of muon energy loss is by delta ray emission and elastic scattering processes.

%% file: chapters/6results.tex
\section{Results}\label{res}
Following the procedures explained in the previous Sections of this work, it has been possible to estimate the values of $Q_\mu = (14080 \pm 50)$ ADC channels, $W_2 = (24.9 \pm 0.4)\%$, $W_1 = (49.61 \pm 0.09)\%$ and $\eta = E_2/E_1 = 2.11 \pm 0.05$. The values of $W_1$, $W_2$ and $\eta$ are extracted from the Geant4 simulation as it has already been optimised for muons in this energy range. The physics in the FLUKA simulation is currently being tuned, but the first results presented in Section \ref{sim} are qualitatively promising. Hence, the value of the calibration factor $k$ in Eq. \ref{eq:flux} is 
\begin{align}\label{eq:k}
    k = (5.87 \pm 0.14) \cdot 10^{-3} (\text{s} \cdot \text{ADC ch.})^{-1}\text{.} 
\end{align}

\begin{figure}[htbp]
    \centering
    \includegraphics[width=0.75\linewidth]{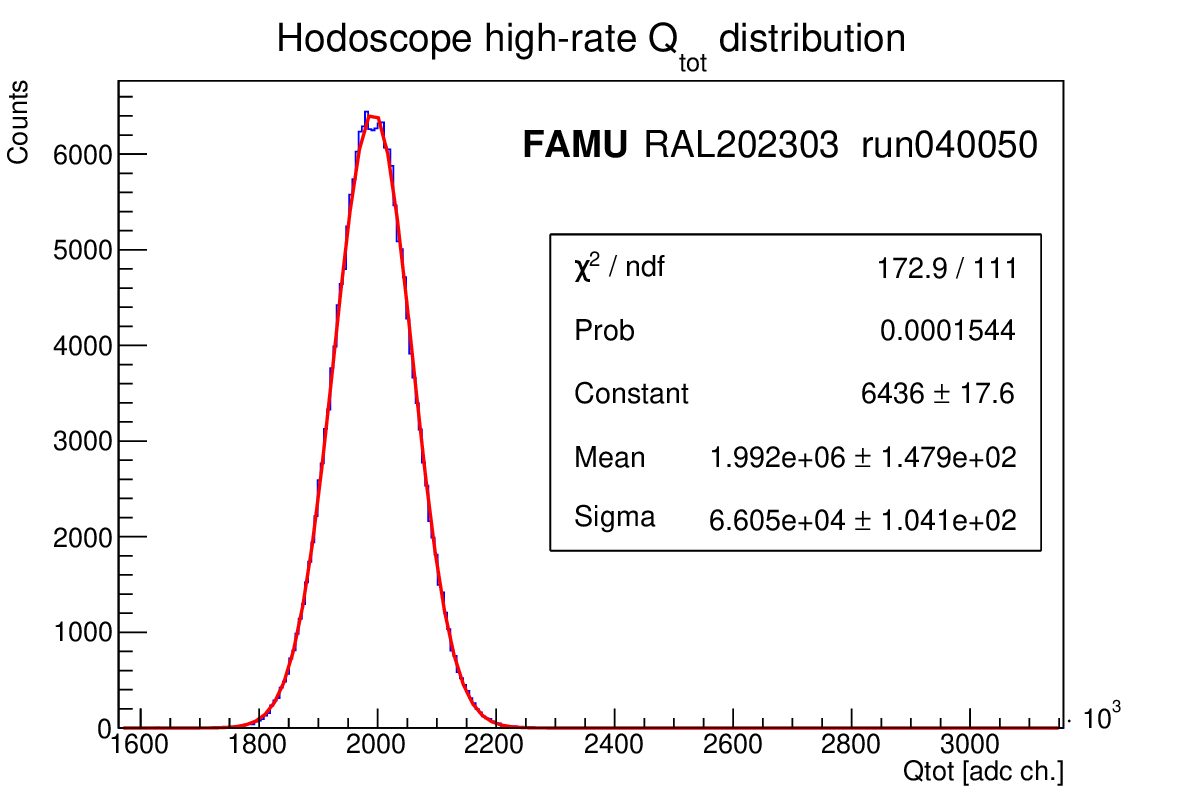}
    \caption{Extraction of average $Q_{tot}$ from the fit of a high-rate $Q_{tot}$ histogram with a Gauss profile.}
    \label{fig:Qtot}
\end{figure}

By fitting with a Gaussian profile the full rate beam histogram of the charge deposited in the hodoscope in each muon spill (see Figure \ref{fig:Qtot}), it is possible to extract the average value of $Q_{tot}$ for the analysed run. However, this flux estimation can also be carried out event-by-event simply taking the punctual value of $Q_{tot}$ and converting it into punctual muon flux.
Taking the mean value and converting it into mean muon rate by applying Eq. \ref{eq:flux} with the value of $k$ in Eq. \ref{eq:k}, one gets $(1.17 \pm 0.03)\cdot 10^4$ muons/s. This value has been obtained with synchrotron current $\sim 85$\% the maximum value. It is consistent with the expected order of magnitude for the 55 MeV/c negative muon flux at full synchrotron current ($> 10^4$ muons/s) (\cite{matsuzaki2001, hillier2019}). 

\begin{figure}[htbp]
    \centering
    \includegraphics[width=0.75\linewidth]{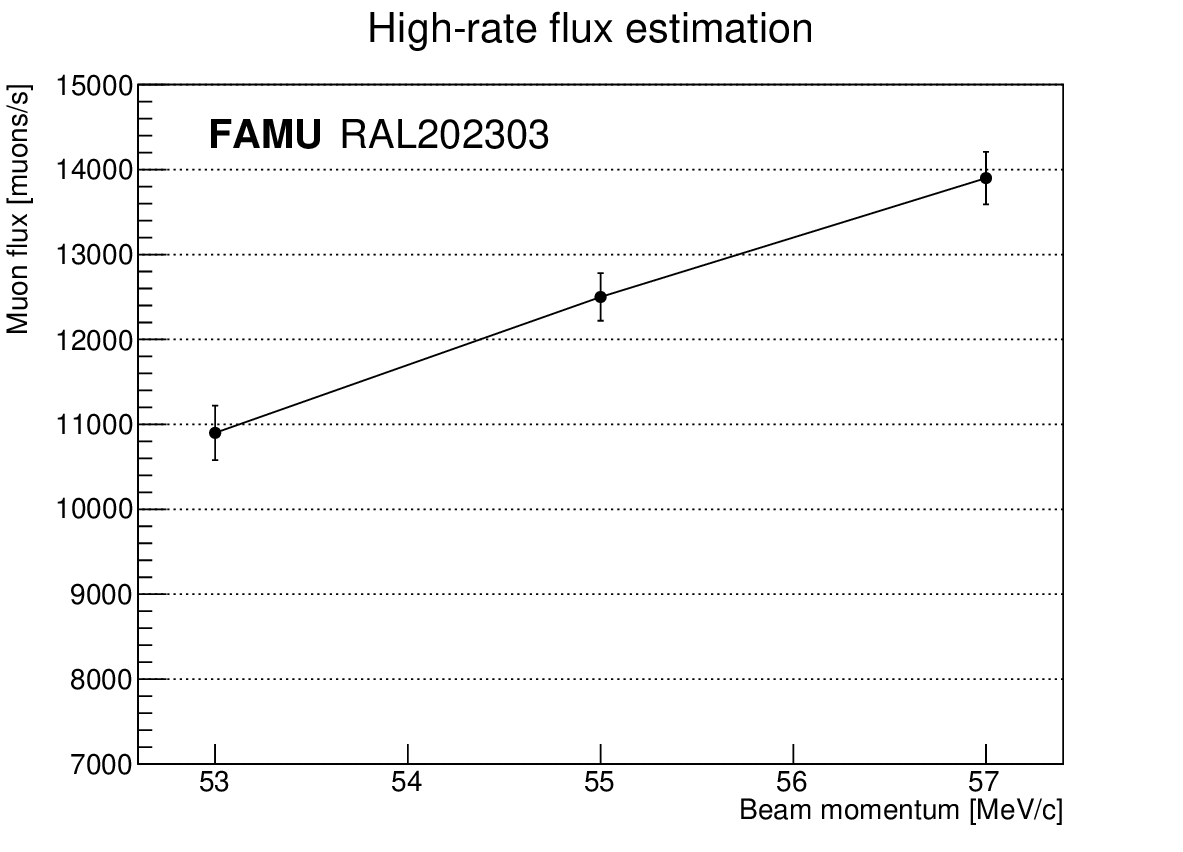}
    \caption{Estimation of muon flux from three values of beam momentum. The correction factor for the value of $Q_{\mu}$ is extracted from the simulation.}
    \label{fig:rate:p}
\end{figure}

By extracting weighing factors for $Q_{\mu}$ at other momenta from the simulation, it has also been possible to estimate the muon flux during high-rate measurements at momentum different from 55 MeV/c. The result is presented in Figure \ref{fig:rate:p} and the trend is increasing with momentum, as expected.

%% file: chapters/7conclusion.tex
\section{Conclusion}\label{con}
A full calibration protocol for a beam hodoscope to be used as flux monitor has been explained, applied and tested. 
While the beam repetition rate $r$ is known, the values of $W_2$, $W_1$ and $\eta$ are extracted from simulation and $Q_\mu$ from low-rate measurements allowing single-particle events. 
In particular, by modeling the detector in a simulation toolkit (Geant4/FLUKA), $W_2$ and $W_1$ are obtained by counting the primary muons interacting with fibres in both planes of the detector or with one fibre only, respectively, whereas $\eta$ is the ratio of the mean deposited energies in the two cases.  On the other hand, $Q_\mu$ is obtained by tuning the beampipe (bending and quadrupole) magnets to deliver a small fraction of beam. By tuning the magnets to allow single-particle spills to reach the detector, it had been possible to determine the amount of charge deposited by muons hitting two fibres by imposing coincidence between the two planes during data analysis. 
This allowed to calculate the calibration constant $k$ to convert the high-rate deposited charge $Q_{tot}$ into muon rate $\varphi_\mu$. The muon rate plays a crucial role in the data normalisation for the FAMU experiment.

This protocol can be applied to other similar detectors in order to use them for the same particle counting task in cases where single-particle discrimination is not possible. 



This procedure can be carried out with any scintillating fibre-based hodoscope to be used as muon beamline monitor, with various applications. For example, Muonic atom X-Ray Spectroscopy ($\upmu$-XES) for elemental and isotopic analysis (\cite{clemenza2019, cataldo, rossini:meteorite}) is a non-destructive technique for the depth-dependent characterisation of materials of interest such as Cultural Heritage samples. It consists in a spectroscopic analysis of the muonic atom X-rays emitted by a sample. Hence, knowing the injected beam rate would give important information about the number of atoms created, helping the quantification of the elements and isotopes present in the sample. The presence of a calibrated beam hodoscope in such applications would therefore help improving the technique.

%% file: frontiers.bbl
\begin{thebibliography}{25}
\providecommand{\natexlab}[1]{#1}
\expandafter\ifx\csname urlstyle\endcsname\relax
  \providecommand{\doi}[1]{doi:\discretionary{}{}{}#1}\else
  \providecommand{\doi}{doi:\discretionary{}{}{}\begingroup \urlstyle{rm}\Url}\fi
\providecommand{\selectlanguage}[1]{\relax}
\providecommand{\bibAnnoteFile}[1]{%
  \IfFileExists{#1}{\begin{quotation}\noindent\textsc{Key:} #1\\
  \textsc{Annotation:}\ \input{#1}\end{quotation}}{}}
\providecommand{\bibAnnote}[2]{%
  \begin{quotation}\noindent\textsc{Key:} #1\\
  \textsc{Annotation:}\ #2\end{quotation}}

\bibitem[{Agostinelli et~al.(2003)Agostinelli, Allison, Amako et~al.}]{GEANT}
Agostinelli, S., Allison, J., Amako, K., et~al. (2003).
\newblock Geant4—a simulation toolkit.
\newblock \emph{NIM A} 506, 250--303.
\newblock \doi{10.1016/S0168-9002(03)01368-8}
\bibAnnoteFile{GEANT}

\bibitem[{Ahdida et~al.(2022)Ahdida, Bozzato, Calzolari et~al.}]{FLUKA2}
Ahdida, C., Bozzato, D., Calzolari, D., et~al. (2022).
\newblock New capabilities of the fluka multi-purpose code.
\newblock \emph{Frontiers in Physics} 9, 788253.
\newblock \doi{10.3389/fphy.2021.788253}
\bibAnnoteFile{FLUKA2}

\bibitem[{Antognini et~al.(2022)Antognini, Hagelstein, and Pascalutsa}]{Antognini}
Antognini, A., Hagelstein, F., and Pascalutsa, V. (2022).
\newblock The proton structure in and out of muonic hydrogen.
\newblock \emph{Annu. Rev. Nucl. Part. Sci} , 389–418\doi{10.1146/annurev-nucl-101920-024709}
\bibAnnoteFile{Antognini}

\bibitem[{Battistoni et~al.(2015)Battistoni, Boehlen, Cerutti et~al.}]{FLUKA1}
Battistoni, G., Boehlen, T., Cerutti, F., et~al. (2015).
\newblock Overview of the fluka code.
\newblock \emph{Annals of Nuclear Energy} 82, 10--18.
\newblock \doi{10.1016/j.anucene.2014.11.007}
\bibAnnoteFile{FLUKA1}

\bibitem[{Bonesini et~al.(2019)Bonesini, Benocci, Bertoni et~al.}]{bonesini2018}
Bonesini, M., Benocci, R., Bertoni, R., et~al. (2019).
\newblock The upgraded beam monitor system of the famu experiment at riken–ral.
\newblock \emph{NIM A} 936, 592--594.
\newblock \doi{10.1016/j.nima.2018.08.092}
\bibAnnoteFile{bonesini2018}

\bibitem[{Bonesini et~al.(2017)Bonesini, Bertoni, Chignoli et~al.}]{bonesini2017}
Bonesini, M., Bertoni, R., Chignoli, F., et~al. (2017).
\newblock The construction of the fiber-sipm beam monitor system of the r484 and r582 experiments at the riken-ral muon facility.
\newblock \emph{Journal of Instrumentation (JINST)} 12, C03035.
\newblock \doi{10.1088/1748-0221/12/03/C03035}
\bibAnnoteFile{bonesini2017}

\bibitem[{Brun and Rademakers(1997)}]{ROOT}
Brun, R. and Rademakers, F. (1997).
\newblock Root — an object oriented data analysis framework.
\newblock \emph{NIM A} 389, 81--86.
\newblock \doi{10.1016/S0168-9002(97)00048-X}
\bibAnnoteFile{ROOT}

\bibitem[{Carbone et~al.(2015)Carbone, Bonesini, Bertoni et~al.}]{carbone:2015}
Carbone, R., Bonesini, M., Bertoni, R., et~al. (2015).
\newblock The fiber-sipmt beam monitor of the r484 experiment at the riken-ral muon facility.
\newblock \emph{Journal of Instrumentation (JINST)} 10, C03007.
\newblock \doi{10.1088/1748-0221/10/03/C03007}
\bibAnnoteFile{carbone:2015}

\bibitem[{Carlson(2015)}]{Carlson}
Carlson, C. (2015).
\newblock The proton radius puzzle.
\newblock \emph{Progress in Particle and Nuclear Physics} 82, 59--77.
\newblock \doi{https://doi.org/10.1016/j.ppnp.2015.01.002}
\bibAnnoteFile{Carlson}

\bibitem[{Cataldo et~al.(2022)Cataldo, Clemenza, Ishida, and Hillier}]{cataldo}
Cataldo, M., Clemenza, M., Ishida, K., and Hillier, A. (2022).
\newblock A novel non-destructive technique for cultural heritage: Depth profiling and elemental analysis underneath the surface with negative muons.
\newblock \emph{A.D. . Appl. Sci.} 12, 4237.
\newblock \doi{10.3390/app12094237}
\bibAnnoteFile{cataldo}

\bibitem[{Clemenza et~al.(2019)Clemenza, Baldazzi, Ballerini et~al.}]{clemenza2019}
Clemenza, M., Baldazzi, G., Ballerini, G., et~al. (2019).
\newblock Chnet-tandem experiment: Use of negative muons at riken-ral port4 for elemental characterization of “nuragic votive ship” samples.
\newblock \emph{NIM A} 936, 27--28.
\newblock \doi{10.1016/j.nima.2018.11.076}
\bibAnnoteFile{clemenza2019}

\bibitem[{Dal~Maso et~al.(2023)Dal~Maso, Barchetti, Francesconi et~al.}]{dalmaso:2023}
Dal~Maso, G., Barchetti, F., Francesconi, M., et~al. (2023).
\newblock Beam monitoring detectors for high intensity muon beams.
\newblock \emph{NIM A} 1047, 167739.
\newblock \doi{/10.1016/j.nima.2022.167739}
\bibAnnoteFile{dalmaso:2023}

\bibitem[{Hillier et~al.(2019)Hillier, Lord, Ishida, and Rogers}]{hillier2019}
Hillier, A.~D., Lord, J.~S., Ishida, K., and Rogers, C. (2019).
\newblock Muons at isis.
\newblock \emph{Philosophical Transactions of the Royal Society A} 377, 20180064.
\newblock \doi{10.1098/rsta.2018.0064}
\bibAnnoteFile{hillier2019}

\bibitem[{Lord et~al.(2011)Lord, McKenzie, Baker et~al.}]{Lord}
Lord, J., McKenzie, I., Baker, P., et~al. (2011).
\newblock Design and commissioning of a high magnetic field muon spin relaxation spectrometer at the isis pulsed neutron and muon source.
\newblock \emph{Rev. Sci. Instrum.} 82, 073904.
\newblock \doi{10.1063/1.3608114}
\bibAnnoteFile{Lord}

\bibitem[{Matsuzaki et~al.(2001)Matsuzaki, Ishida, Nagamine et~al.}]{matsuzaki2001}
Matsuzaki, T., Ishida, K., Nagamine, K., et~al. (2001).
\newblock The riken-ral pulsed muon facility.
\newblock \emph{NIM A} 465, 365--383.
\newblock \doi{10.1016/S0168-9002(01)00694-5}
\bibAnnoteFile{matsuzaki2001}

\bibitem[{Papa et~al.(2015)Papa, Barchetti, Gray, Ripiccini, and Rutar}]{papa:2015}
Papa, A., Barchetti, F., Gray, F., Ripiccini, E., and Rutar, G. (2015).
\newblock A multi-purposed detector with silicon photomultiplier readout of scintillating fibers.
\newblock \emph{NIM A} 787, 130--133.
\newblock \doi{10.1016/j.nima.2014.11.074}
\bibAnnoteFile{papa:2015}

\bibitem[{Papa et~al.(2019)Papa, Rutar, Barchetti, Hildebrandt, and Kettle}]{papa:2019}
Papa, A., Rutar, G., Barchetti, F., Hildebrandt, M., and Kettle, P. (2019).
\newblock A fast and quasi non-invasive muon beam monitor working at the intensity frontier.
\newblock \emph{NIM A} 936, 634--635.
\newblock \doi{10.1016/j.nima.2018.10.145}
\bibAnnoteFile{papa:2019}

\bibitem[{Pizzolotto et~al.(2020)Pizzolotto, Adamczak, Bakalov et~al.}]{pizzolotto2020}
Pizzolotto, C., Adamczak, A., Bakalov, F., D., et~al. (2020).
\newblock The famu experiment: muonic hydrogen high precision spectroscopy studies.
\newblock \emph{Eur. Phys. J. A} 56, 185.
\newblock \doi{10.1140/epja/s10050-020-00195-9}
\bibAnnoteFile{pizzolotto2020}

\bibitem[{Rossini et~al.(2024{\natexlab{a}})Rossini, Adamczak, Bakalov et~al.}]{rossiniFAMU:2023setup}
Rossini, R., Adamczak, A., Bakalov, D., et~al. (2024{\natexlab{a}}).
\newblock Status of the detector setup for the famu experiment at riken-ral for a precision measurement of the zemach radius of the proton in muonic hydrogen.
\newblock \emph{Journal of Instrumentation (JINST)} 19, C02034.
\newblock \doi{10.1088/1748-0221/19/02/C02034}
\bibAnnoteFile{rossiniFAMU:2023setup}

\bibitem[{Rossini et~al.(2023{\natexlab{a}})Rossini, Benocci, Bertoni et~al.}]{rossini2023_2}
Rossini, R., Benocci, R., Bertoni, R., et~al. (2023{\natexlab{a}}).
\newblock Characterisation of a scintillating fibre-based hodoscope exposed to the cnao low-energy proton beam.
\newblock \emph{NIM A} 1046, 167746.
\newblock \doi{10.1016/j.nima.2022.167746}
\bibAnnoteFile{rossini2023_2}

\bibitem[{Rossini et~al.(2023{\natexlab{b}})Rossini, Benocci, Bertoni et~al.}]{rossini2023_1}
Rossini, R., Benocci, R., Bertoni, R., et~al. (2023{\natexlab{b}}).
\newblock Characterisation of muon and proton beam monitors based on scintillating fibres with a sipm read-out.
\newblock \emph{NIM A} 1046, 167684.
\newblock \doi{10.1016/j.nima.2022.167684}
\bibAnnoteFile{rossini2023_1}

\bibitem[{Rossini et~al.(2024{\natexlab{b}})Rossini, Benocci, Bertoni et~al.}]{rossiniHodoJINST}
Rossini, R., Benocci, R., Bertoni, R., et~al. (2024{\natexlab{b}}).
\newblock Characterisation of a low-momentum high-rate muon beam monitor for the famu experiment at the cnao-xpr beam facility.
\newblock \emph{Journal of Instrumentation (JINST)} 19, C01024.
\newblock \doi{10.1088/1748-0221/19/01/C01024}
\bibAnnoteFile{rossiniHodoJINST}

\bibitem[{Rossini et~al.(2023{\natexlab{c}})Rossini, Di~Martino, Agoro et~al.}]{rossini:meteorite}
Rossini, R., Di~Martino, D., Agoro, T., et~al. (2023{\natexlab{c}}).
\newblock A new multidisciplinary non-destructive protocol for the analysis of stony meteorites: gamma spectroscopy, neutron and muon techniques supported by raman microscopy and sem-eds.
\newblock \emph{J. Anal. At. Spectrom.} 38, 293--302.
\newblock \doi{10.1039/D2JA00263A}
\bibAnnoteFile{rossini:meteorite}

\bibitem[{Vacchi et~al.(2023)Vacchi, Mocchiutti, Adamczak et~al.}]{Vacchi}
Vacchi, A., Mocchiutti, E., Adamczak, A., et~al. (2023).
\newblock Investigating the proton structure: The famu experiment.
\newblock \emph{Nuclear Physics News} 33, 9--16.
\newblock \doi{10.1080/10619127.2023.2198913}
\bibAnnoteFile{Vacchi}

\bibitem[{Vlachoudis(2009)}]{FLAIR}
Vlachoudis, V. (2009).
\newblock {Flair: A powerful but user friendly graphical interface for FLUKA}.
\newblock In \emph{{International Conference on Mathematics, Computational Methods \& Reactor Physics 2009}}. 790--800
\bibAnnoteFile{FLAIR}

\end{thebibliography}
